# Correlated Paramagnetism and Interplay of Magnetic and Phononic Degrees of Freedom in 3$d$-5$d$ Coupled La$_2$CuIrO$_6$

Birender Singh[1], Deepu Kumar[1], Kaustuv Manna[2], A. K. Bera[3], G. Aslan Cansever[2], A. Maljuk[2], S. Wurmehl[2,4,] B. Büchner[2,4] and Pradeep Kumar[1*]

[1]*School of Basic Sciences, Indian Institute of Technology Mandi, Mandi-175005, India*

[2]*Leibniz-Institute for Solid State and Materials Research, (IFW)-Dresden, D-01171 Dresden, Germany*

[3]*Solid State Physics Division, Bhabha Atomic Research Centre, Mumbai 400 085, India*

[4]*Institute of Solid State Physics, TU Dresden, 01069 Dresden, Germany*

*email id: pkumar@iitmandi.ac.in

**Abstract:** Conventional Paramagnetism - a state with finite magnetic moment per ion sans long range magnetic ordering, but with lowering temperature the moment on each ion picks up a particular direction, breaking rotational symmetry, and results into long-range magnetic ordering. However, in systems with competing multiple degrees of freedom this conventional notion may easily break and results into short range correlation much above the global magnetic transition temperature. La$_2$CuIrO$_6$ with complex interplay of spins ($s$ =1/2) on Cu site and pseudo-spin ($j$ =1/2) on Ir site owing to strong spin-orbit coupling provides fertile ground to observe such correlated phenomena. By a comprehensive temperature dependent Raman study, we have shown the presence of such a correlated paramagnetic state in La$_2$CuIrO$_6$ much above the long-range magnetic ordering temperature ($T_N$). Our observation of strong interactions of phonons, associated with Cu/Ir octahedra, with underlying magnetic degrees of freedom mirrored in the observed Fano asymmetry, which remarkably persists as high as ~ 3.5$T_N$ clearly signals the existence of correlated paramagnetism hence broken rotational symmetry. Our detailed analysis also reveals anomalous changes in the self-energy parameters of the phonon modes, i.e. mode frequencies and linewidth, below $T_N$, providing a useful gauge for monitoring the strong coupling between phonons and magnetic degrees of freedom.

## INTRODUCTION

Materials with *5d* transition metals have attracted numerous research activities due to their rich magnetic quantum states of matter. In particular, the *5d* transition metal oxides provide fertile ground to explore interplay between spin-orbit coupling (SOC) and electron-electron correlation, which are within comparable energy scale, resulting in interesting phenomena such as $J_{eff} = 1/2$ Mott-insulator, quantum-spin liquids, Weyl semimetals, axion insulators, unconventional superconductors and realization of the Kitaev model with spin-spin correlation [1-8]. In the current studies of double perovskite iridates and iridium based oxides i.e. $Na_4Ir_3O_8$, $Sr_2IrO_4$, and $Na_2IrO_3$, in +4 ($Ir^{4+}$, $5d^5$) and +5 ($Ir^{5+}$, $5d^4$) oxidation state [9-16] shows Mott-insulating ground state with effective total angular momentum $J_{eff} = 1/2$ driven by strong SOC. Despite of intense research on double perovskite iridates, the factors facilitating the rich magnetic ground state are still under debate. Recently, the focus is on double perovskites with co-existence of *3d* and *5d* transition metals, such as $La_2RIrO_6$ (R = Zn, Mn, Ni, Fe, Co and Cu) [17-23], which have provided new frontier in unrevealing the complex quantum state of these materials owing to competing nature of multiple degrees of freedom. Due to the compact nature of *3d*-orbitals, they have strong electronic correlation and highly quenched orbital angular momentum due to presence of non-degeneracy of *d*-orbitals originated from the Jahn-Teller effect, however, in contrast spatially extended *5d*-orbitals exhibit opposite behaviour i.e. strongly pronounced SOC and weak electronic correlation. Also, in case of 3*d* transition metal oxides, the pseudo-spin degrees of freedom are active owing to very weak SOC and as a result Jahn-Teller effect generate strong coupling between phonons and pseudo-spin degrees of freedom [24]. On the other hand, for *5d* systems, strong SOC is expected to quench pseudo-spin dynamics. However, it has been suggested in recent studies [25-28] that even in case of *5d* system Jahn-Teller mechanism is active despite strong SOC and couple the lattice and pseudo-spin degrees of freedom. It has also been described that at moderate inter-site hopping by electrons, SOC is not strong enough to quench pseudo-spin dynamics completely and this results into emergence of non-zero moments attributed to the entanglement between spin and pseudo-spin degrees of freedom mediated by lattice [26-27]. Therefore, it becomes pertinent to probe the role of lattice degrees of freedom in these systems to understand their role in governing the underlying mechanism responsible for their rich magnetic ground state.

The double perovskite $La_2CuIrO_6$ possess Cu in +2 (i.e. $Cu^{2+}$ ($3d^9$); spin, s = 1/2) and Ir in +4 (i.e. $Ir^{4+}$ ($5d^5$); pseudo-spin, j=1/2) oxidation state. Interestingly, spins are active in Cu sub-lattice and pseudo-spins in Ir sub-lattice and with lowering temperature intricate coupling between them gives rise to the long range Antiferromagnetic (AFM) ordering at $T_N$ ~ 74 K. It is observed that short range spin-spin correlation starts emerging within the $Cu^{2+}$ sub-lattice at temperature as high as ~ 113 K. At the same time there are some weak signature of correlation seen till temperature as high as ~ 180 K, indicating weak correlation within the paramagnetic (PM) phase [23]. Earlier studies have focused on the role of SOC, on-site interaction, inter-site hopping and probing their complex interplay [25-27], however, a detailed study deciphering the role of lattice degrees of freedom is lacking on this system. In the present work we have undertaken such a study using inelastic light scattering via Raman spectroscopy as a function of temperature. Raman scattering is a very powerful tool to probe the spin, pseudo-spin, lattice and electronic degrees of freedom and have played a very important role in our current understanding of the perovskite system [29-36]. Therefore, it is natural that Raman scattering studies may be very crucial for uncovering the entanglement of lattice degrees of freedom with magnetic and/or electronic degrees of freedom in the double perovskite $La_2CuIrO_6$, where the ground state is expected to be driven by complex interplay of SOC, on-site interaction, etc. From our experimental observation, we found clear signature of renormalization of phonon modes below $T_N$ attributed to the strong coupling of phonons with the magnetic degrees of freedom. Interestingly, anomalous decrease in phonon lifetime is observed below magnetic ordering, clearly reflecting phonon interaction with other quasi-particle excitation. In addition, Fano-asymmetric profile is observed in one of the prominent phonon mode at low temperature, signifying the strong coupling of lattice vibrations with the underlying magnetic degrees of freedom. Remarkably, this coupling survives at a temperature much higher than $T_N$ in the PM phase reflecting the correlated nature of the magnetic moments much above the long range magnetic ordering temperature.

## EXPERIMENTAL AND COMPUTATIONAL DETAILS

The single phase polycrystalline samples of $La_2CuIrO_6$ were prepared as described in Ref. 23. Unpolarized Raman spectra were excited with 532 nm solid-state laser in backscattering geometry using Raman spectrometer (Labram HR-Evolution). The excitation laser line was focused on sample surface through 50x magnification long working distance objective with a numerical aperture of 0.8. Laser power was kept very low < 1mW to avoid local heating on the sample surface. The Raman shifted light was analysed via spectrometer with 1800 lines/mm grating coupled with a Peltier cooled charged-coupled detector (CCD). The temperature dependent measurement was done from 5-330 K using closed-cycle He-flow cryostat (Montana) in a cryogenically pumped vacuum with ± 0.1 K or better temperature accuracy.

The structural optimization and the zone-centred phonon frequencies calculations were performed utilizing plane-wave approach as implemented in QUANTUM ESPRESSO [37]. We have used linear response approach within density functional perturbation theory (DFPT) to calculate the dynamical matrix and zone-centred phonon frequencies. The ultra-soft pseudopotentials (PBESol) is used as an exchange-correlation functional with plane-wave energy cutoff 25 Ry and the charge density cutoff of 250 Ry. The Monkhorst-pack scheme with 4 x 4 x 4 *k*-point mesh is used for numerical integration of Brillouin zone (BZ). The phonon frequencies at zone center were calculated with experimental lattice parameters as $a = 5.8000$ Å, $b = 7.7279$ Å, $c = 5.6023$ Å [23].

## RESULTS AND DISCUSSION
### A. Raman Scattering and Zone Centre Phonon Calculations

La$_2$CuIrO$_6$ crystallizes in triclinic structure ($P\bar{1}$; space group No. 2) [23]. Within triclinic phase La$_2$CuIrO$_6$ exhibits twenty-four Raman active modes, at the gamma point in Brillouin zone, in the irreducible representation given as $24A_g$. In three-dimensional structure La$_2$CuIrO$_6$ consists of (Cu/IrO$_6$) octahedra with two apical and four equatorial oxygen atoms. Figure 1 shows the unpolarized Raman spectra of La$_2$CuIrO$_6$ in a wide spectral range of 60-800 cm$^{-1}$ at 5 K. Spectra are fitted with a sum of Lorentzian functions in the spectral range of 60-610 cm$^{-1}$ and with a Fano function in 610-800 cm$^{-1}$ to attain comprehensive quantitative information about mode frequency, linewidth and intensity (the individual modes are shown by thin blue lines, and the resultant fit is shown by a thick red line). The observed spectra comprise twelve modes labelled as S1-S12 (see table-I) and based on our first principle calculations we have assigned these modes to be first order phonon modes. The low frequency modes (< 300 cm$^{-1}$) mainly originate due to the vibration of La/Cu/Ir ions, while the modes in high frequency range ($300 < \omega < 800$ cm$^{-1}$) arises due to the stretching and bending vibration of oxygen-atoms associated with Cu/IrO$_6$ octahedra.

To gain microscopic insight about the behavior of phonon vibrations in La$_2$CuIrO$_6$, we performed zone-centred phonon calculations via DFPT. The DFPT calculated mode frequencies using experimental lattice parameters are in very good agreement with the experimentally observed frequencies at 5 K (see table-I). A comparison between calculated phonon frequency with the experimental lattice parameters and experimentally observed frequency at 5 K has been made as: $|\bar{\omega}_r|_{\%} = \frac{100}{N} \sum_i \left|\frac{\omega_i^{cal} - \omega_i^{exp}}{\omega_i^{exp}}\right|$, where $i = 1, 2 ----, N$ is the number of Raman active modes (12 here) and $|\bar{\omega}_r|_{\%}$ is the average of the absolute relative differences in percentage. The estimated relative average in the mode frequencies is observed to be only ~ 0.2 percent. Figure 2 shows the calculated eigen vector for the phonon modes S1, S3, S6, S8, S11 and S12 where, black arrow indicates the direction of lattice vibration associated with respective atom within the unit cell structure and the length of arrow indicate the magnitude of lattice vibration.

### B. Temperature Dependence of the Phonon Modes

Figure 3 (a-c) illustrate the temperature dependence of mode frequencies and full widths at half maximum (FWHM) of the prominent phonon modes. Following observation can be made: (i) Significant anharmonic effect is observed above magnetic ordering temperature ($T_N$ ~ 74 K ) for all the observed phonon modes, remarkably for S1, S3, S4, S5, S10 and S11 as 4.3, 4.9, 2.8, 2.9, 2.4 and 2.1 % respectively; however, it is ~ 1.2 % for other observed modes (i.e. S2, S6, S7, S8 and S9). (ii) Interestingly, below $T_N$ majority of the modes show anomalous softening till 40 K; however below ~ 40 K phonon frequencies become nearly constant. The temperature independent behaviour of phonon frequencies at low temperature (< 40 K ) may be due to the change in magnetic spin-structure and/or emergence of weak Ferromagnetic (FM) ordering as suggested in earlier transport studies [23]. The observed phonon softening below $T_N$ is found to be more pronounced for low energy phonon modes S1, S2, S3, S4 and S5 as 0.9, 0.3, 0.4, 0.7 and 0.4 % respectively, while for higher energy phonon modes (i.e. S6, S7, S8, S9, S10 and S11) it is ~ 0.2-0.3 % . We note that anomaly in the mode frequencies for different modes occurs at slightly different temperature, see inset of Fig. 3 (a, b). This variation in the peak may arise from the fact that different phonon modes involve vibration of different atoms with varying magnitude as well as direction ( please see Fig. 2 describing eigen vector for different modes ) and this may result into dynamic modulation of different magnetic exchange paths and therefore spin correlation effect on phonons may vary. This may also result from some finite contribution from spin fluctuation due to quantum and thermal effects [38]. We note that the exact spin configuration in the ground state is still not clear [23], we hope that our experimental results on the different degrees of renormalization of different phonons modes may be a good starting point to qualitatively understand the behaviour of phonons as a function of different spin configurations, and may be crucial in deciphering the exact spin configuration in the ground state. (iii) The temperature dependence of line-width shows normal behaviour (i.e. line-width narrowing with decreasing temperature) above $T_N$, however below $T_N$, line-width increases significantly with decreasing temperature till lowest recorded temperature as shown in Figure 3c for S1, S3, S6 and S7 phonon modes. The observed increase in line-width below $T_N$ is as high as 20 % with respect to its value at $T_N$ ( see Figure 3c ). Conventionally, in solids linewidth decreases (i.e. phonon lifetime increases) with decreasing temperature due to reduced anharmonicity at low temperature. As anharmonic effect is more pronounced at high temperature, due to large atomic movement from their mean position, resulting in a large phonon-phonon interaction, which is reflected as the decrease in the phonon life-time, however, at low temperature influence of thermal fluctuation is significantly quenched due to decrease in phonons interaction and as a result phonon life-time increases or FWHM decreases. The observed line broadening upon entering into the magnetically ordered phase clearly indicate that anharmonicity alone is not sufficient to capture the full *T*-dependent behaviour suggesting that some additional decay channel becomes active owing to long range correlation between spin/ pseudo-spin degrees of freedom.

To understand the microscopic origin of the temperature evolution of the phonon modes above $T_N$, one may assume as atoms in a crystal lattice vibrating about their equilibrium position at all temperature. Potential, U(r), that an atom experience may be expressed as: $U(r) = U(r_0) + r \frac{\partial U}{\partial r}\big|_{r=r_0} + r^2 \frac{\partial^2 U}{\partial r^2}\big|_{r=r_0} + r^3 \frac{\partial^3 U}{\partial r^3}\big|_{r=r_0} + ---$, where the first term is some constant and may be set to zero for convenience, second term will go to zero as we are expanding about an equilibrium position and the third term is pure harmonic term and will not give any temperature dependent effect on the phonons [39]. Therefore, cubic and higher order terms are solely responsible for the temperature dependent behaviour and hence are responsible for an anharmonic behaviour of a phonon mode. Anharmonic potential responsible for the temperature dependence of a mode may be written as $U_{anh}(r) = gr^3 + mr^4 + ---- \equiv \beta(a^+a^+a + a^+aa) + \gamma(a^+a^+a^+a + a^+a^+aa) + ----$, where $a^+$, $a$ are the phonon creation and annihilation operator, respectively. Cubic term ($gr^3$) will give change in frequency and linewidth as $\Delta\omega = \omega(T) - \omega_0 = A(1 + \frac{2}{e^x-1})$, $\Delta\Gamma = \Gamma(T) - \Gamma_0 = C(1 + \frac{2}{e^x-1})$, respectively, where, $x = \frac{\hbar\omega_0}{2k_BT}$, which involve decay of an optical phonon into two phonons of equal frequency. Quartic term ($mr^4$) involve decay of an optical phonon into three phonons of equal frequencies and the corresponding change in the frequency and linewidth as $\Delta\omega = \omega(T) - \omega_0 = B(1 + \frac{3}{e^y-1} + \frac{3}{(e^y-1)^2})$, $\Delta\Gamma = \Gamma(T) - \Gamma_0 = D(1 + \frac{3}{e^y-1} + \frac{3}{(e^y-1)^2})$, respectively, where $y = \frac{\hbar\omega_0}{3k_BT}$ [40]. Constant coefficient, A/B and C/D, describe the strength of phonon-phonon interaction involving three/four phonon process, respectively. In principle one may extend this procedure, however for low temperature range three phonon process is dominating and four phonons contribution is small, and other higher order process start contributing only at much higher temperature. Considering both three and four phonon process, we have fitted the mode frequency $\omega(T)$ and line width $\Gamma(T)$ using the following functions [40]:

$$\omega(T) = \omega_0 + A(1+\tfrac{2}{e^x-1}) + B(1+\tfrac{3}{e^y-1}+\tfrac{3}{(e^y-1)^2})$$

and

$$\Gamma(T) = \Gamma_0 + C(1+\tfrac{2}{e^x-1}) + D(1+\tfrac{3}{e^y-1}+\tfrac{3}{(e^y-1)^2})$$

respectively, where $\omega_0$ and $\Gamma_0$ is the mode frequency and line-width at zero temperature. As the long range magnetic ordering appears below ~ 74 K, so we have fitted the mode frequencies and line-width only in the temperature range of ~ 80 to 330 K. Fitting parameters are listed in the Table-I. We note that for majority of modes constant B/D is smaller as compared to A/C, respectively, suggesting that three phonon process is dominating but there is finite contribution from four phonon process.

The temperature evolution of all phonon modes is in good agreement with the phonon-phonon anharmonic interaction model in higher temperature regime ( 80-330 K ) (see red colour solid lines above 80 K in Fig. 3 and 5 (a)). However, in the lower temperature region (< 80 K) most of the phonon modes shows strong renormalization (see Figure 3 (a and b) and 5 (a)) indicating that the phonon anharmonic interaction model cannot capture the complete picture to explain the observed anomalies. Similarly, the phonon linewidth $\Gamma(T)$ does not follow the simple phonon anharmonic model below $T_N$ i.e. increase in line-width (decrease in phonon life-time) below $T_N$ indicates the relative sensitivity of phonon life-time with magnetic ordering (see Figure 3 (c)). Our clear observation of anomalous increase in line-width below $T_N$ suggests strong interaction of phonons with magnetic degrees of freedom, which leads to decrease in their lifetime. Within this scenario, the origin of anomalous temperature dependence may be attributed to strong coupling of phonons with the underlying magnetic degrees of freedom.

The phonon renormalization due to coupling of magnetic degrees of freedom with lattice dynamics through the modulation of exchange integral may be understood in the following way [34]. The crystal potential (U) due to displacement of an ion from its equilibrium position within the harmonic approximation is given as, $U(r) = U_0 + \sum_{i, i \neq j} J_{ij}(r) \vec{S_i}.\vec{S_j}$, where $U_0 = \frac{1}{2}kr^2$, $k$ and $r$ are the force constant and displacement from equilibrium position, respectively, $J_{ij}$ is the exchange energy constant and $\vec{S_i}.\vec{S_j}$ is spin-spin interaction between $i^{th}$ and $j^{th}$ magnetic ion. The second derivative of crystal potential is, $\frac{\partial^2 U}{\partial r^2} = k + \sum_{i, i \neq j} \frac{\partial^2 J_{ij}}{\partial r^2} <\vec{S_i}.\vec{S_j}>$ where, first term represents the harmonic force constant and the corresponding second part arises due to modulation of exchange coupling owing to lattice vibration. The renormalization of the phonon mode frequency due to coupling with magnetic degrees of freedom arises due to this additional contribution. Therefore, frequency shift below $T_N$ is in direct relation with spin-phonon coupling, $\Delta\omega_{sp-ph} = \lambda <\vec{S_i}.\vec{S_j}>$, where $\lambda = \frac{\partial^2 J_{ij}}{\partial r^2}$ is spin-phonon coupling coefficient and it can be positive or negative [35, 41-42] depending on the phonon hardening and softening below $T_N$. Microscopically, sign of $\lambda$ will be govern by the detailed nature of $J_{ij}$ and the symmetry of the phonon modes involved. Furthermore, nearly constant value of phonon frequency in low temperature regime ( < 40 K) suggests the spin-reorientation and this might be due to an incoherent multi-magnetic excitation and/or quantum excitation i.e. appearance of weak intrinsic ferromagnetic component in $La_2CuIrO_6$, which is observed to be prominent in low temperature regime [23]. It is suggested that AFM ordering below $T_N$ is accompanied by residual transverse moments and with further lowering temperature below ~ 40-50 K, these transverse moments gives rise to weak ferromagnetism. Based on the proposed spin configuration for the ground state of $La_2CuIrO_6$ system [23], the change in mode frequency say for the mode S1, S6 and S12 may be written as a function of exchange parameter as, $\Delta\omega^{S1}_{sp-ph} = \{\frac{1}{\mu_O}[\frac{\partial^2 J^{AF}_{1xz}}{\partial r^2_{O_{3,4}}} + \frac{\partial^2 J^{FM}_{2xz}}{\partial r^2_{O_{1,2}}} + \frac{\partial^2 J^{FM}_{2y}}{\partial r^2_{O_{5,6}}}] + \frac{1}{\mu_{cu/Ir}}[\frac{\partial^2 J^{AF}_{1xz}}{\partial r^2_{cu/Ir}} + \frac{\partial^2 J^{FM}_{2xz}}{\partial r^2_{cu/Ir}}]\}(M^2(T)/M^2_{max})$

$\Delta\omega^{S6}_{sp-ph} = \{\frac{1}{\mu_O}[\frac{\partial^2 J^{FM}_{2xz}}{\partial r^2_{O_{1,2}}} + \frac{\partial^2 J^{FM}_{2y}}{\partial r^2_{O_{5,6}}}] + \frac{1}{\mu_{cu}}[\frac{\partial^2 J^{FM}_{2xz}}{\partial r^2_{cu}}]\}(M^2(T)/M^2_{max})$ and $\Delta\omega^{S12}_{sp-ph} = \frac{1}{\mu_O}[\frac{\partial^2 J^{AF}_{1xz}}{\partial r^2_{O_{3,4}}} + \frac{\partial^2 J^{FM}_{2y}}{\partial r^2_{O_{5,6}}}](M^2(T)/M^2_{max})$,

respectively. Where $\mu_{O,Cu,Ir}$ is the effective mass of oxygen, Cu and Ir atoms involved in the mode vibrations. $J^{AF}_1$ and $J^{FM}_2$ are the Antiferromagnetic and Ferromagnetic exchange parameter between Cu and Ir and $O_{1-6}$ represents different oxygen atoms in the Cu/Ir octahedra. From these above relations it is clear that different phonon modes will be renormalized in different way as the exchange parameters involved are different for different modes in addition to the fact that magnitude of vibration of atoms involved is also different for different modes (see Fig. 2). Hence, we attribute the observed phonon anomalies to strong hybridization of magnetic excitation and lattice dynamics through spin-phonon coupling.

Microscopically, renormalization of the phonon frequency due to spin ordering below the magnetic transition temperature can be correlated with the magnetization and is given as [34], $\Delta\omega_{sp-ph} \propto <\vec{S_i}.\vec{S_j}>$, where $<\vec{S_i}.\vec{S_j}>$ is proportional to $\frac{M^2(T)}{M^2_{max}}$, $\Delta\omega_{sp-ph} = [\omega(T) - \omega_{5K}] \propto (\frac{M^2(T)}{M^2_{max}})$ where, $\omega_{5K}$ is mode frequency at lowest recorded temperature (5 K), $M(T)$ is the magnetization and $M_{max}$ corresponds to the saturation magnetization at applied dc magnetic field ($H_{dc}$) of 10 kOe [23]. Note that the frequency shift $\Delta\omega_{sp-ph}$, for the prominent mode S12, is in very good agreement with the $\frac{M^2(T)}{M^2_{max}}$ curve (see Figure 4), suggesting that the anomalous softening indeed arises from the strong coupling of phonons with the magnetic degrees of freedom.

One of the prominent mode (S12) near ~ 83 meV ( ~ 670 cm$^{-1}$ ) shows Fano asymmetry. This asymmetry, also known as Fano resonance, in the line-shape reflects the coupling of a quantized excitation, phonon here, with a broad continuum of typically electronic and/or magnetic origin. We note that we observed asymmetry only for one of the prominent mode i.e. S12. As Fano resonance has its roots in strong coupling of phonons to a continuum of excitation, effectively involving the coupling of two quasi-particle excitations. So, a phonon mode of similar energy and symmetry as that of underlying continuum will be affected most as these two parameters decides the nature of coupling [43], this may be the reason why we do see this asymmetry only in one of the observed phonon modes. In literature Fano asymmetry has been reported for different systems, interestingly out of the total numbers of observed modes very few modes show Fano asymmetry [28, 30, 44-46]. The observed Fano-asymmetric phonon mode (S12) was fitted well by a Fano function [43] (see Figure 1) described as $F(\omega) = I_0 [\frac{(q+\varepsilon)^2}{(1+\varepsilon^2)}]$, where $\varepsilon = (\frac{\omega - \omega'}{\Gamma})$, $\omega'$ and $\Gamma$ are the phonon frequency and linewidth, respectively, and $q$ define the nature of asymmetry. Figure 5 (a-c) display the temperature dependence of the mode frequency ($\omega'$), line-width ($\Gamma$), Fano asymmetry parameter (1/|q|) and the intensity; and the phase diagram determined via Raman scattering is shown in Figure 5 (d). Temperature dependence of the mode frequency and line-width shows similar trend as that of other phonon modes i.e. phonon softening and broadening in line width (~ 5.1 %) below $T_N$. The phonon mode softening below $T_N$ is prominent down to ~ 40 K and thereafter it is nearly constant down to 5 K. Interestingly, the Fano asymmetry parameter (1/|q|) shows strong temperature dependence (see Figure 5 (b)), it is nearly constant below $T_N$ and

continuously decrease with increasing temperature till ~ 250 K, thereafter it remains constant up to 330 K. The observed maximum asymmetry in line-shape below $T_N$ suggests the strong coupling between underlying magnetic degrees of freedom with the phonons.

Furthermore, it is interesting to note that the intensity profile also exhibits anomalous temperature dependence as the intensity of this asymmetric phonon mode remain constant up to $T_N$ on heating and sharp decrease in the intensity is observed in the temperature range of 80 to ~120 K, thereafter, slow decrease is seen till ~ 250 K and remains nearly constant after that. This observation of anomalous temperature dependence of intensity i.e. sharp increase in intensity below 120 K and constant profile below $T_N$ correlates very well with the proposed emergence of short-range correlation with in the $Cu^{2+}$ sublattice below ~ 113 K [23]. In the earlier study it has been suggested that short-range spin-spin correlation within the $Cu^{2+}$ (spin, $s = 1/2$) start developing below 113 K and on further decrease in temperature $Ir^{4+}$ (pseudo-spin, $j = 1/2$) sublattice come into the picture and strong entanglement between these two sublattice is observed below ~ 74 K leading to AFM transition [23]. As Raman is a very sensitive technique, any modification in the Raman tensor due to interaction with other quasi-particles will be reflected in the observed spectra. Therefore, the inevitably sharp increase in intensity of the Fano-asymmetric mode can be attribute to the development of short range spin-spin correlation of $Cu^{2+}$ below ~120 K.

In systems with strong pseudo-spin/spin-phonon coupling, lattice degrees of freedom provide an extra probe of magnetic excitation. The observation of Fano asymmetry is a clear signature of coupling with the underlying magnetic degrees of freedom. The continuous build-up of the Fano asymmetry with decreasing temperature after ~ 250 K clearly signals the emergence of spin-spin and pseudo-spin correlation within the PM phase. Interestingly, this asymmetry is significant till very high temperature ( ~ $3.5T_N$ ) indicating the presence of correlated or quasi PM phase. It has been suggested that Mott-insulator driven by strong SOC provide fertile ground to be precursor for quantum spin liquids phase/correlated PM/Quasi PM. Within this correlated paramagnetic phase the long range ordering is suppressed but the incoherent local spin ordering continue to survive, which decay to zero in a non-exponential fashion against the exponential fall for the long range order phenomena. We note that similar signature of quantum/correlated paramagnetism have been reported for other Ir-based systems and two-leg spin ladder system [30, 44] as well. Also, for the case of $La_2CuIrO_6$ only ~ 25 % of the expected magnetic entropy is recovered across the magnetic transition, implying that majority of the entropy is released above $T_N$ signalling building up of the short-range correlation much above long range magnetic ordering temperature [23]. Our observation of these strong anomalies in the phonon self-energy parameters and intensity profile clearly suggest the emergence of finite correlation within the PM phase well above the long range magnetic ordering temperature.

## Summary and Conclusions

We performed detailed phonon dynamics study of double perovskite $La_2CuIrO_6$ using temperature dependent Raman scattering along with the first principle DFT based calculations. The anomalous softening of the phonon frequencies and line-width increase in the long range magnetic ordered phase is attributed to the strong coupling of phonons and magnetic degrees of freedom. The appearance of Fano-asymmetric line shape of the prominent mode near ~ 670 cm$^{-1}$, which survive till very high temperature (~ $3.5T_N$ ), suggests the presence of the correlated paramagnetic phase much above $T_N$. Density functional perturbation theory based calculated phonon frequencies at the zone-centre with the experimental lattice parameters were found to be in very good agreement with experimental observed values at 5 K. Our results cast a crucial light on the role of lattice degrees of freedom in this *3d-5d* hybrid system and suggests that lattice degrees of freedom should be treated at par with other relevant degrees of freedom driving the ground state of these systems and calls for further exploring the correlated PM phase in these *3d-5d* coupled systems.

**Acknowledgment**

PK thanks the Department of Science and Technology, India, for the grant under INSPIRE Faculty scheme and Advanced Material Research Center, IIT Mandi, for the experimental facilities. The authors at Dresden thanks Deutsche Forschungsgemeinschaft (DFG) program for financial support through SFB 1143 (Project B01), and Emmy Noether Grant No. WU595/3-3.

**Table-I:** List of the experimentally observed phonon frequencies at 5 K and fitting parameters (fitted using equations as described in the text) and calculated phonon frequencies using experimental lattice parameters. Units are in cm$^{-1}$. Mode assignment (with displacement of atoms involved) has been done on the basis of our first principle calculations.

| Mode Assignment | Exp. $\omega$ 5 K | Fitted Parameters | | | | | | DFT $\omega$ |
|---|---|---|---|---|---|---|---|---|
| | | $\omega_0$ | A | B | $\Gamma_0$ | C | D | |
| S1-$A_g$ (La,Cu,Ir,O) | 95.1 | 96.8 ± 0.2 | -0.62 ± 0.2 | 0.002 ± 0.001 | 6.13 ± 0.4 | 0.25 ± 0.1 | -0.001 ± 0.001 | 91.3 |
| S2-$A_g$ (La,O) | 136.9 | 136.7 ± 0.2 | 0.10 ± 0.03 | -0.003 ± 0.001 | - | - | | 136.8 |
| S3-$A_g$ (La,Cu,Ir,O) | 154.1 | 156.9 ± 0.8 | -1.78 ± 0.4 | 0.002 ± 0.001 | 6.76 ± 0.6 | 1.24 ± 0.3 | -0.006 ± 0.001 | 151.3 |
| S4-$A_g$ (La,Cu,O) | 280.2 | 284.1 ± 1.2 | -2.37 ± 0.5 | -0.04 ± 0.02 | - | - | | 282.2 |
| S5-$A_g$ (Cu,O) | 315.9 | 321.9 ± 2.1 | -3.95 ± 0.9 | -0.07 ± 0.03 | - | - | | 323.1 |
| S6-$A_g$ (Cu,O) | 379.2 | 382.9 ± 1.8 | -4.27 ± 1.3 | -0.08 ± 0.05 | 12.01 ± 2.1 | 9.34 ± 2.6 | -0.20 ± 0.2 | 382.4 |
| S7-$A_g$ (Cu,O) | 397.2 | 401.4 ± 2.1 | -4.51 ± 1.0 | 0.07 ± 0.04 | 17.39 ± 2.3 | -2.21 ± 1.2 | 0.43 ± 0.18 | 401.7 |
| S8-$A_g$ (O) | 445.1 | 443.0 ± 1.6 | -3.29 ± 1.6 | -0.73 ± 0.14 | - | - | | 447.1 |
| S9-$A_g$ (Cu,O) | 497.4 | 493.9 ± 2.1 | 4.59 ± 0.18 | -0.86 ± 0.04 | - | - | | 493.1 |
| S10-$A_g$ (Cu,O) | 528.0 | 544.1 ± 1.2 | -13.92 ± 0.3 | 0.22 ± 0.08 | - | - | | 541.1 |
| S11-$A_g$ (O) | 584.5 | 600.4 ± 1.1 | -15.38 ± 0.4 | 0.13 ± 0.09 | - | - | | 570.8 |
| S12-$A_g$ (O) | 669.7 | 679.2 ± 0.5 | -9.63 ± 0.2 | 0.23 ± 0.07 | 6.34 ± 0.9 | 8.13 ± 0.3 | 0.009 ± 0.3 | 659.1 |

**FIGURE CAPTION:**

**FIGURE 1:** (Color online) Raman spectra of La$_2$CuIrO$_6$ at 5 K. Solid thin lines are the fit of individual modes with a sum of Lorentzian functions, and solid thick red line shows the total fit to the experimental data. Figure on right show the Fano-fit for mode S12, red line is the fitted curve and blue spheres represent raw data.

**FIGURE 2:** (Color online) Schematic showing the eigen vectors for some of the prominent phonon modes. Green, blue, dark yellow and red spheres represent La, Cu, Ir and oxygen atoms, respectively. Black arrows on the atoms represent the direction of motion and the size of the arrow represent the magnitude of vibration. *a*, *b* and *c* represent the crystallographic axis.

**FIGURE 3:** (Color online) (a and b) Temperature evolution of the phonon frequencies of modes S1, S3, S4, S5, S6, S7, S10 and S11 (Inset: shows zoomed view below 125 K and solid green line is guide to eye); (c) Temperature evolution of linewidth of the few prominent modes S1, S3, S6 and S7. Solid red lines are the fitted curve as described in the text and solid black lines and dotted red lines are the guide to eye.

**FIGURE 4:** (Color online) Temperature-dependence of $\Delta\omega_{sp-ph}$, for the prominent phonon mode S12, and $\frac{M^2(T)}{M^2_{max}}$ (magnetization at $H_{dc}$ = 10 kOe). Broad yellow line is guide to the eye.

**FIGURE 5:** (Color online) (a) Temperature-dependence of the mode frequency and linewidth (solid line is the fitted curve as described in the text) of the mode S12 (~ 670 cm$^{-1}$); Inset: shows zoomed view below 125 K and solid green line is guide to eye. (b) Temperature evolution of the Fano asymmetry parameter 1/|q| (c) Normalized intensity with solid red lines as guide to the eye, inset shows temperature evolution of mode S12. (d) Phase diagram as determined by the Raman scattering.

**Figure 1:**

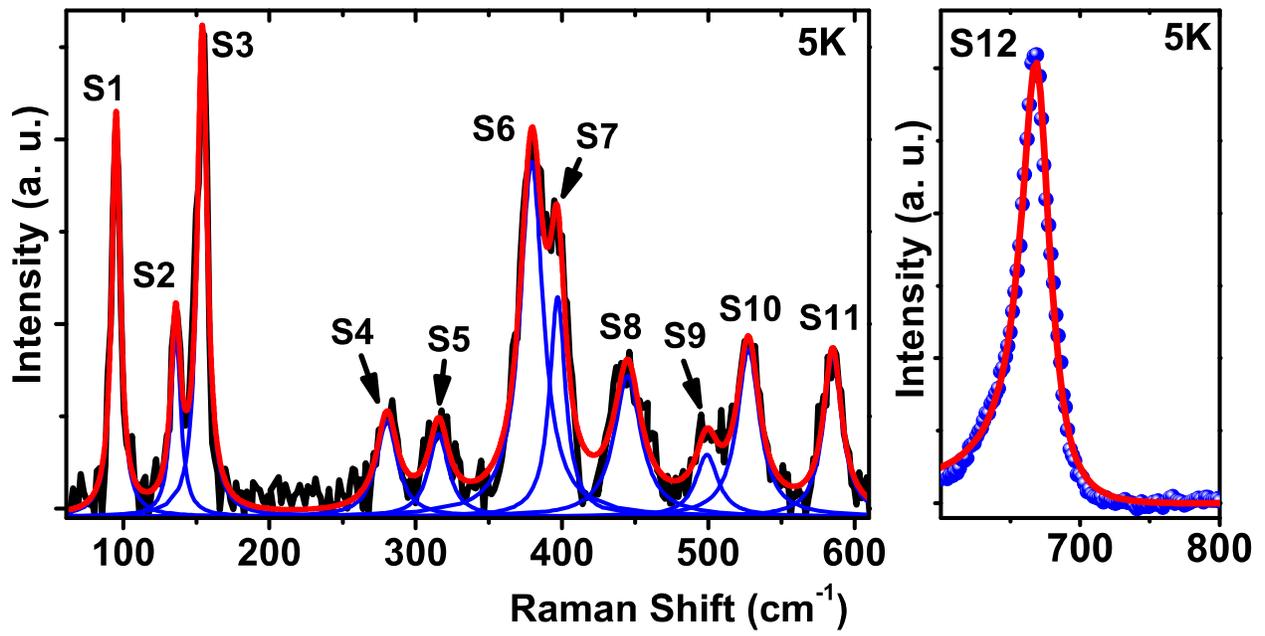

**Figure 2:**

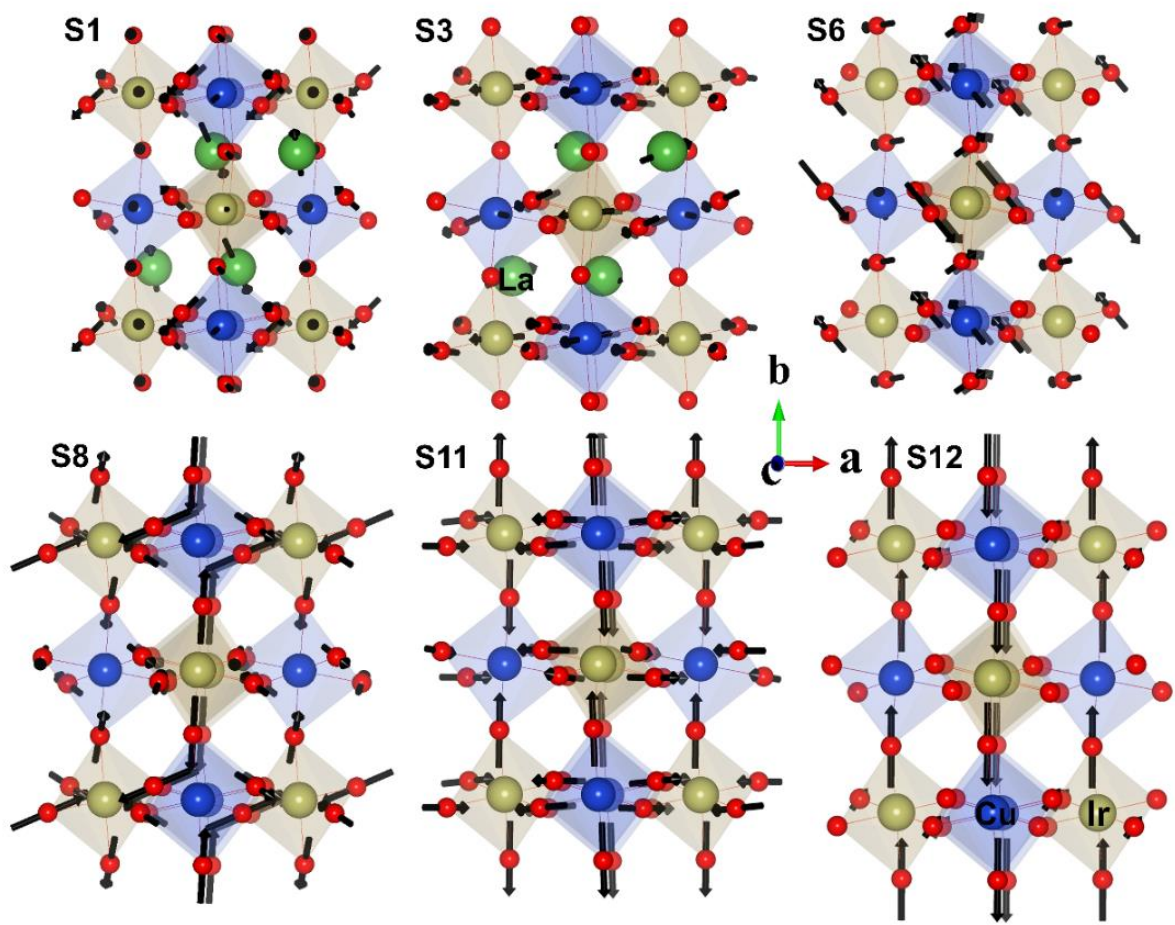

**Figure 3:**

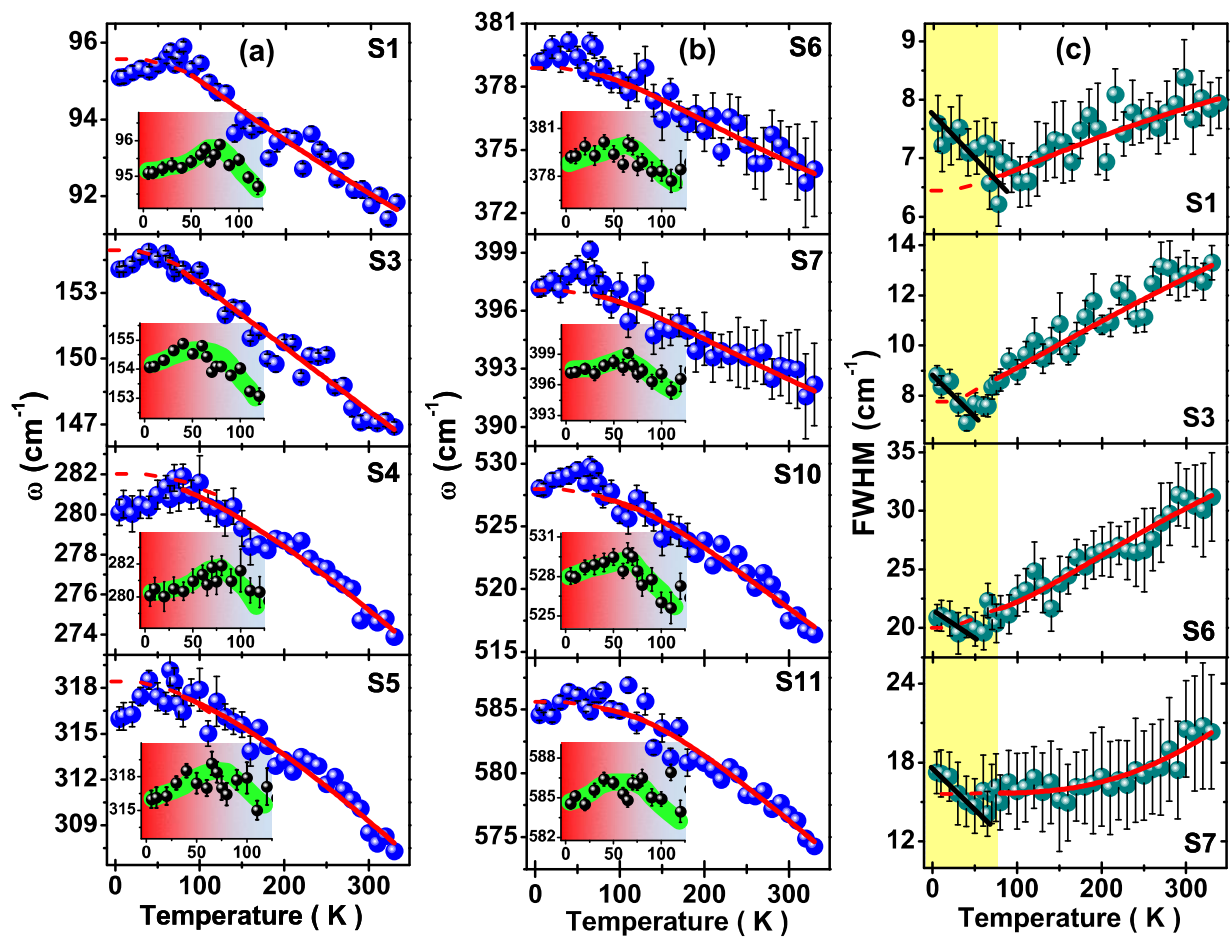

**Figure 4:**

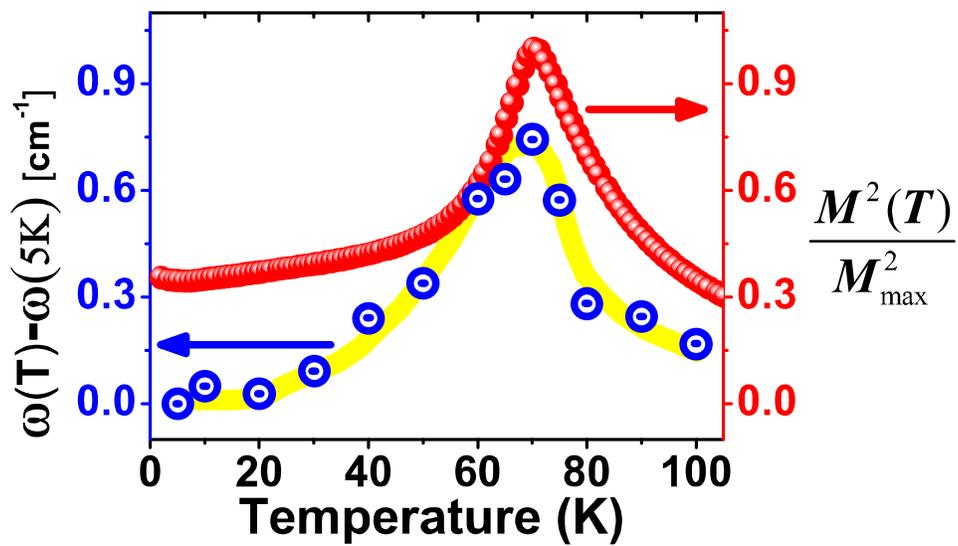

**Figure 5:**

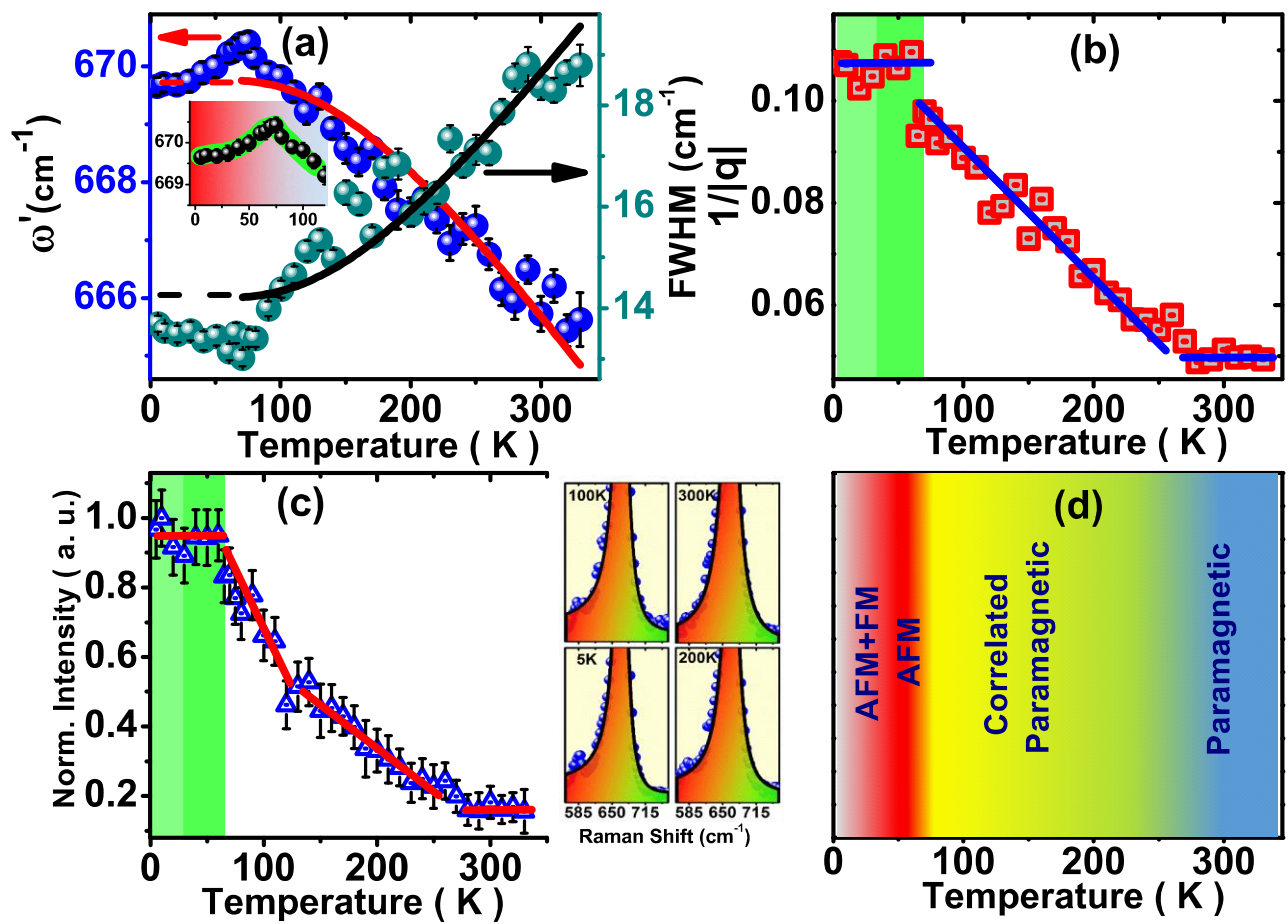